\documentclass[11pt]{article}
\usepackage{hyperref}
\pdfoutput=1
\usepackage{natbib}
\begin{document}
\title{An Integrated Study of Vortex Formation of Freely Flying Insects}
\author{Hui Wan, Yan Ren, Zongxian Liang, Zach Gaston, Haibo Dong\\
        \vspace{6pt} Department of Materials and Mechanical Engineering, \\
	Wright State University, Dayton, OH 45435, USA}
\maketitle

\begin{abstract}
This is a short introduction illustrating movies submitted to "Fluid Dynamics Videos".
\end{abstract}
\section{Introduction}
An integrated approach is introduced to study the free flight of insects (e.g. dragonflies)(Dong et al. 2010).
First, High-Speed Photogrammetry is used to record various flight motions. As an example, a dragonfly in a backward taking-off is shown in the video.
Three-dimensional surface reconstruction techniques are then applied to obtain the data of body trajectory, wing kinematics and deformation.
Based on these data, direct numerical simulation (DNS) of full body is conducted with our in-house high fidelity CFD solver, which is based on sharp interface immerse boundary method (Mittal et al., 2008). The vortex structure created in the process of dragonfly taking-off can be clearly seen.
As a further step, tracers and Lagrangian coherent structure (Shadden et al. 2005) are used to understand the vortex formation and facilitate the vortex identification.
\\
%
%

\hspace{-0.5cm}\textbf{Reference}

\begin{enumerate}
\item
Dong, H. and Koehler, C. and Liang, Z. and Wan, H. and Gaston, Z., ``An integrated analysis of a dragonfly in free flight", 40th AIAA Fluid Dynamics Conference and Exhibit, AIAA 2010-4390
\item
Mittal, R. and Dong, H. and Bozkurttas, M. and Najjar, F.M.and Vargas, A. and von Loebbecke, A., ``A versatile sharp interface immersed boundary method for incompressible flows with complex boundaries",
Journal of Computational Physics, 2008, Vol. 227, pp. 4825-4852
\item
Shadden, S.C., Lekien, F., Marsden, J.E.,
``Definition and properties of Lagrangian coherent structures from finite-time Lyapunov exponents in two-dimensional aperiodic flows",
Physica D: nonlinear Phenomena, 2005, Vol. 212, pp. 271-304
\end{enumerate}

%
\end{document}